\documentclass[
nofootinbib,
 amsmath,amssymb,
 aps,
pra,
twocolumn
]{revtex4-1}

\usepackage{dcolumn}
\usepackage{bm}
\usepackage{epsfig}
\usepackage{graphicx}
\usepackage{amsfonts}
\usepackage[figuresright]{rotating}
\usepackage{psfrag}
\usepackage{float} 
\usepackage{footmisc}

\input epsf.sty

\newcommand{\be}{\begin{equation}}
\newcommand{\ee}{\end{equation}}
\newcommand{\ba}{\begin{array}}
\newcommand{\ea}{\end{array}}
\newcommand{\bqa}{\begin{eqnarray}}
\newcommand{\eqa}{\end{eqnarray}}

\begin{document}

\title{Photonic de Haas-van Alphen effect}
\author{Kejie Fang}
\altaffiliation[Current address: ]{Thomas J. Watson, Sr., Laboratory of Applied Physics, California Institute of Technology, Pasadena, CA 91125}
\homepage{https://sites.google.com/site/bacbever/}
\affiliation{Department of Physics, Stanford University, Stanford,
California 94305, USA}
\author{Zongfu Yu}
\affiliation{Department of Electrical Engineering, Stanford
University, Stanford, California 94305, USA}
\author{Shanhui Fan}
\affiliation{Department of Electrical Engineering, Stanford
University, Stanford, California 94305, USA}

\begin{abstract}

Based on the recently proposed concept of effective gauge potential and magnetic field for photons, we numerically demonstrate a photonic de Haas-van Alphen effect. We show that in a dynamically modulated photonic resonator lattice exhibiting an effect magnetic field, the trajectories of the light beam at a given frequency have the same shape as the constant energy contour for the photonic band structure of the lattice in the absence of the effective magnetic field. 

\end{abstract}
\pacs{}

\maketitle

\section{Introduction}

The use of externally-imposed electric and magnetic fields is of crucial importance in controlling both the classical and quantum motions of electrons. It will be of practical and fundamental importance to explore similar mechanisms for controlling the flow of photons. While an effective electric field for photons can be straightforwardly created with the use of spatially-inhomogeneous dielectric or metallic structures \cite{blochosc0,russell,blochosc1,blochosc2,blochosc3,zener,kivshar, lieven}, creating an effective magnetic field for photons has been more elusive. 

Very recently, it was pointed out \cite{our1,our2,our3} that an effective magnetic field for photons can emerge in a dynamic system undergoing temporal modulation. In these systems the phase of the modulations correspond to a gauge potential for photons \cite{our1,our3}. And hence with a spatially inhomogeneous distribution of modulation phases, an effective magnetic field for photons can emerge \cite{our2}. Since the temporal modulation can break time-reversal symmetry \cite{yu}, such an effective magnetic field also breaks time-reversal symmetry, in contrast to some of the recent related proposals to create a gauge field for photons based on a spin degree of freedom for photons where time-reversal symmetry is not broken \cite{hafezi, umu,photonicTI,strain}. 

Ref. \cite{our2} showed that a photon in the presence of a uniform effective magnetic field experiences an effective Lorentz force. In this paper, we consider the interplay between the effective magnetic field and the photonic band structure. We show that for a photon in a dynamic resonator lattice exhibiting an effective magnetic field, its motion in fact exhibits a photonic analogue of the electronic de Hass-van Alphen effect, with the circular trajectory as seen in the Lorentz force demonstrated in Ref. \cite{our2} being only a special example of such photonic de Hass-van Alphen effect. 

The paper is organized as follows. In Section II, we briefly review the method to create effective magnetic field for photons and related numerical simulation method. In Section III, we numerically demonstrate a photonic de Haas-van Alphen effect, where a light beam propagating under an effective magnetic field traces out a trajectory with a shape that corresponds to the constant energy contour of the underlying photonic resonator lattice. In Section IV, we conclude by discussing the experimental requirement to realize these novel effects.

\section{Model system, theoretical background, and numerical methods}
In this section, we discuss our model system of a dynamically modulated photonic resonator lattice. We briefly review the mechanism to generate an effective gauge field and magnetic field for photons in such dynamically modulated lattice \cite{our1,our2}. We also provide a brief discussion of the numerical simulation methods that we use.

\subsection{Model Hamiltonian and Floquet Bandstructure}
\begin{figure}[H]
\centering\epsfig{file=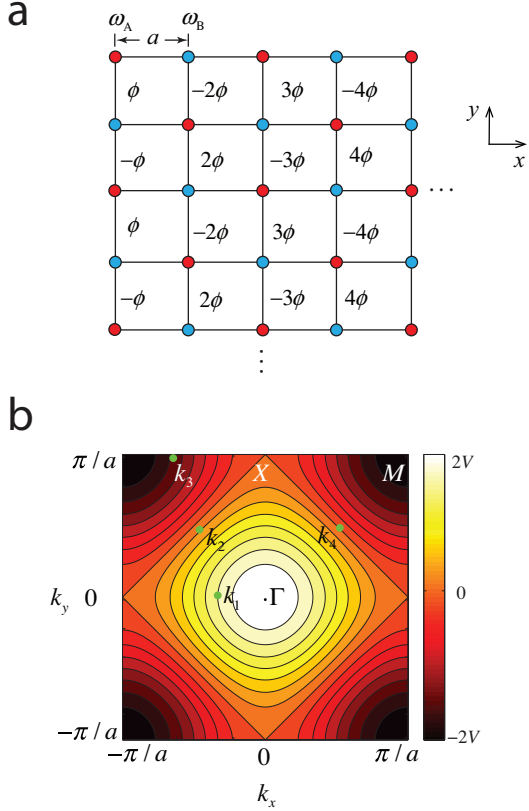,clip=1,width=0.8\linewidth,angle=0}\vspace{0.3cm}\hspace{0.3cm}
\caption{(Color online). \textbf{a} Schematic of a photonic crystal resonator lattice with dynamically modulated nearest-neighbor coupling. The modulation phase is zero along the $x$ direcition, and varies in space along the $y$ direction as indicated in the figure. \textbf{b} Constant energy contours in the first Brillouin zone of a square lattice with lattice constant $a$ and nearest-neighbor coupling strength $V$ in the absence of effective magnetic field, which corresponds to the case with $\phi = 0$ everywhere in a.  }
\label{lattice}
\end{figure}

Our model system consists of a two-dimensional photonic resonator lattice as shown in Fig. \ref{lattice}a. The lattice has a square unit cell and each unit cell contains two resonators $A$ and $B$ with different resonant frequencies $\omega_A$ and $\omega_B$ ($\omega_A>\omega_B$), respectively. We assume only nearest-neighbor coupling with a form of $V\textrm{cos}(\Omega t+\phi)$, where $V$ is the coupling strength, $\Omega$ and $\phi$ are the modulation frequency and phase respectively. The dynamics of the fields on this lattice is then described by the coupled mode equation 
\be\label{schrodinger}
 i\frac{d}{dt}|\psi\rangle=H(t)|\psi\rangle,
\ee
where $|\psi\rangle$ is the photon amplitude. The Hamiltonian $H(t)$ of this resonator lattice is \bqa\label{Ht} H(t)&=&\omega_A\sum\limits_i a^\dagger_ia_i+\omega_B\sum\limits_j b^\dagger_jb_j\\\nonumber &&+\sum\limits_{\langle ij \rangle}V\textrm{cos}(\Omega t+\phi_{ij})(a^\dagger_ib_j+b^\dagger_ja_i), \eqa where $a_i^\dagger (a_i)$ and $b_j^\dagger (b_j)$ are the creation (annihilation) operators of the $A$ and $B$ resonators, respectively, and $\phi_{ij}$ is the phase of the modulation between resonators at site $i$ and $j$. 

Because $H(t+2\pi/\Omega)=H(t)$, the solution of Eq. \eqref{schrodinger} can be written as $|\psi(t)\rangle=e^{-i\epsilon t}|\chi(t)\rangle$, where $|\chi(t+2\pi/\Omega)\rangle=|\chi(t)\rangle$, and $\epsilon \, (\textrm{mod} \, \Omega)$ is the \emph{quasi-energy} \cite{sherley,samba}. Using Eq. \eqref{schrodinger}, we see that $|\chi(t)\rangle$ satisfy an eigenvalue equation: 
\be
\label{eig}(i\partial_t-H(t))|\chi(t)\rangle=-\epsilon|\chi(t)\rangle,
\ee
where the left-most minus sign is put in for later convenience. Eq. \eqref{eig} can be solved by a Fourier expansion,
\bqa
\label{f1}|\chi(t)\rangle=\sum\limits_{n=-\infty}^\infty |\chi_n\rangle e^{in\Omega t},\\
\label{f2}H(t)=H_0+H_1e^{i\Omega t}+H_{-1}e^{-i\Omega t}.
\eqa
Substitute Eqs. \eqref{f1} and \eqref{f2} into Eq. \eqref{eig}, and compare the coefficient of $n$th Fourier component, we obtain for all integer $n$
\be\label{coeff}
(H_0-\epsilon+n\Omega)|\chi_n\rangle+H_1|\chi_{n-1}\rangle+H_{-1}|\chi_{n+1}\rangle=0.
\ee
We see if $\epsilon$ is a solution of Eq. \eqref{coeff}, then $\epsilon+m\Omega$ is also a solution for any integer $m$, so we restrict $\epsilon$ to the irreducible zone between $-\Omega/2$ and $\Omega/2$.

When $\phi_{ij}\equiv \textrm{const.}$, the lattice has spatial periodicity, and thus $H(t)$ has good quantum numbers (momenta) $k_x$ and $k_y$. $\epsilon$ as a function of $k_x$ and $k_y$ is the Floquet band structure. 

\subsection{Rotating wave approximation and effective gauge field}
If the modulation is on resonance, i.e. $\Omega=\omega_A-\omega_B$, and the modulation strength satisfies rotating wave approximation $V\ll\Omega$, One can ignore the counter rotating term in Eq. \eqref{Ht}. As a result
\bqa\label{Ht2}
H(t)&\approx&\omega_A\sum\limits_i a^\dagger_ia_i+\omega_B\sum\limits_j b^\dagger_jb_j\\\nonumber &&+\sum\limits_{\langle ij \rangle}[\frac{V}{2}e^{-i(\Omega t+\phi_{ij})}a^\dagger_ib_j+\frac{V}{2}e^{i(\Omega t+\phi_{ij})}b^\dagger_ja_i].
\eqa\label{rwa1}
After transferring to a rotating frame, $a_i(b_j)\rightarrow U_{ii(jj)}c_i(c_j)$ with $U_{ii(jj)}=e^{i\omega_{A(B)}t}$, the Hamiltonian in Eq. \eqref{Ht2} simplifies to 
\bqa \nonumber H_{\rm rwa}&=&UHU^{-1}+i\frac{dU}{dt}U^{-1}\\ &=&\sum\limits_{\langle ij \rangle}\frac{V}{2}(e^{-i\phi_{ij}}c^\dagger_ic_j+e^{i\phi_{ij}}c^\dagger_jc_i). \label{rwa1}\eqa Note in the first equality we have written in matrix form. Eq. \eqref{rwa1} resembles the Hamiltonian of electrons in a lattice under a gauge field, with a Peierls substitution \cite{luttinger}. The effective gauge potential for photons can thus be defined as \be\label{lineint} \int_i^{j} \vec A_{\rm eff}\cdot d\vec l=\phi_{ij} .\ee If the integral of the effective gauge potential around a plaqutte in the lattice is non-zero, then there is an effective magnetic flux through the plaquette. The magnetic field strength is \be B_{\rm eff}=\frac{1}{a^2}\oint \vec A_{\rm eff}\cdot d\vec l, \ee  where $a$ is the distance between two nearest-neighbor resonators. Since the modulation phase distribution can in principle be arbitrarily chosen, there are great flexibilities in specifying different effective magnetic field and gauge potential distribution, as we will exploit in this paper. 

Note under rotating wave approximation and on-resonance condition, the quasi-energy $\epsilon$ of $H(t)$ (solution of Eq. \eqref{coeff}  for the Hamiltonian of Eq. \eqref{Ht2}) becomes the eigen-energy $\epsilon_{\rm rwa}$ of $H_{\rm rwa}$ of Eq. \eqref{rwa1}. This can be proved as follows. We write $|\chi_n\rangle=|\chi_{n,A}\rangle+|\chi_{n,B}\rangle$, where $|\chi_{n,A(B)}\rangle$ is the component in resonators $A(B)$. Under rotating wave approximation, $H(t)$ is given by Eq. \eqref{Ht2}, and thus Eq. \eqref{coeff} becomes 
\bqa
\label{aa}(H_{0,A}-\epsilon+n\Omega)|\chi_{n,A}\rangle+H_{-1}|\chi_{n+1,B}\rangle=0,\\
\label{bb}(H_{0,B}-\epsilon+n\Omega)|\chi_{n,B}\rangle+H_1|\chi_{n-1,A}\rangle=0,
\eqa
where $H_{0,A(B)}=\omega_{A(B)}\sum\limits_{i(j)}a^\dagger_{i(j)}a_{i(j)}$. Replacing $n$ in Eq. \eqref{bb} with $n+1$ and combining these two equations, we obtain an eigenvalue equation:
\begin{widetext}
\bqa\label{eigAB}
\left( \ba{cc}   H_{0,A}+n\Omega & H_{-1} \\ H_1 & H_{0,B}+(n+1)\Omega \ea \right) \left(  \ba{c}  |\chi_{n,A}\rangle \\ |\chi_{n+1,B}\rangle \ea\right)=\epsilon \left(  \ba{c}  |\chi_{n,A}\rangle \\ |\chi_{n+1,B}\rangle \ea\right).
\eqa
\end{widetext}
Since in the matrix form $H_{0,A}+n\Omega=H_{0,B}+(n+1)\Omega$ under $\Omega=\omega_A-\omega_B$, the matrix in Eq. \eqref{eigAB} has the same form as the matrix form of the Hamiltonian $H_\textrm{rwa}$. It follows therefore that $ \epsilon= \epsilon_\textrm{rwa}$.

Thus, since we assume rotating wave approximation through out this paper, we will only consider the band structure of $H_{\rm rwa}$ for simplicity. As an example, Fig. \ref{lattice}b shows the band structure of Eq. \eqref{rwa1} for $\phi_{ij}\equiv 0$, given by 
\be
\epsilon_{\rm rwa}(k_x,k_y)=V(\textrm{cos}(ak_x)+\textrm{cos}(ak_y)).
\ee

\subsection{Numerical methods}
We will numerically simulate the propagation of photon beams in the dynamically modulated resonator lattices, using the time-dependent Hamiltonian of Eq. (\ref{Ht}). We will compare the results of such simulations with theoretical derivations based on the time-independent Hamitonian of Eq. (\ref{rwa1}) which is simpler. We only consider the weak effective magnetic field case, i.e. $a^2B_{\textrm{eff}}\ll 1$. The motion of the photon state is then simulated using the coupled mode equation in the presence of a source \be\label{sch} i\frac{d|\psi\rangle}{dt}=H(t)|\psi\rangle+|s\rangle,\ee where $H(t)$ is of Eq. \eqref{Ht} and $|\psi\rangle=[\sum\limits_{i} v_i(t)a^\dagger_i+\sum\limits_{j}v_j(t)b^\dagger_j]|0\rangle$ is the photon state and $v_{i(j)}(t)$ gives the amplitude at site $i(j)$. The beams are excited by a continuous wave source with a spatial Gaussian profile of the form 
\begin{widetext}
\be\label{source} |s\rangle=\theta(t-t_0)\sum\limits_{x,y}e^{-((x-x_0)^2+(y-y_0)^2)/w^2}e^{i(k_{x0}x+k_{y0}y)-i(\omega_{x,y}+\epsilon_0)(t-t_0)}a^\dagger(b^\dagger)_{\{x,y\}}|0\rangle, \ee 
\end{widetext}
where $w$ is the width of the source, $\{x_0,y_0\}$ is the center of the source, $\{k_{x0},k_{y0}\}$ are the momentum of the beam, $\omega_{x,y}$ is the frequency of the resonator at coordinate $\{x,y\}$, $\epsilon_0$ is determined by the energy band $\epsilon_0=\epsilon(k_{x0},k_{y0})$ of the lattice without effective magnetic field, $t_0$ is the excitation time and $\theta(t)$ is the Heaviside step function. 

We solve Eq. \eqref{sch} using a second-order finite-difference time-domain method \cite{numsch}. We discretize time into a sequence $\{t_n\}$, and $|\psi(t)\rangle$ is acquired through iterations: \be
|\psi(t_{n+1})\rangle=|\psi(t_{n-1})\rangle-2iH(t_{n})|\psi(t_{n})\rangle\delta t-2i|s(t_{n})\rangle\delta t. \ee

From the photon state $|\psi\rangle$ we define the photon beam intensity $\langle\psi | \psi\rangle$. As is important for practical implementation, we will prove that the trajectory of photons as characterized by the beam intensity is independent of the excitation time of the source ($t_0$ in Eq. \eqref{source}) in the long evolution time limit. To assure a well-defined trajectory, we require the Gaussian source to satisfy $w\gg a$. We separate Eq. \eqref{source} into two parts $|s\rangle=|s_A\rangle+|s_B\rangle$, where $|s_{A(B)}\rangle$ has non-vanishing coefficients only in resonator $A(B)$.  First of all, we numerically observed that the two sources $|s_A\rangle$ and $|s_B\rangle$, excited at $t_0=0$, generate photon states that in the long evolution time limit are only different up to a phase, i.e. $|\psi_{A(B)}\rangle=e^{i\alpha_{A(B)}}|\chi\rangle$, and thus the photon beam amplitude distribution $\langle\psi_{A(B)}|\psi_{A(B)}\rangle$ in the long evolution time limit is the same. Based on this, sources of the form $|\tilde s\rangle=e^{i\alpha}|s_A\rangle+e^{i\beta}|s_B\rangle$ excited at $t_0=0$ leads to a same beam trajectory, where $\alpha$ and $\beta$ are two arbitrary phases. Next, we consider an excitation source $|s\rangle$ (Eq. \eqref{source}) with $t_0\neq 0$. At $t=t_0$, the modulation phase for the bond between sites $i$ and $j$ has the form $\phi_{ij} + \Omega t_0$, where $\phi_{ij}$ is the phase at $t=0$ as shown in Fig. \ref{lattice}a. By changing the origin of the time axis from 0 to $t_0$, which corresponds to a transformation of $t \rightarrow t-t_0$ in both the Hamiltonian and the source, the modulation phase distribution at $t=t_0$ for the Hamiltonian becomes the same as that in Fig. \ref{lattice}a, while the source is transformed to $|s^\prime\rangle=e^{i\omega_At_0}|s_A\rangle+e^{i\omega_Bt_0}|s_B\rangle$. As noted above, this source induces the same beam propagation for any $t_0$ under the phase distribution of Fig. \ref{lattice}a. On the other hand, since the physics of the problem does not depend on the detailed choice of the time origin, we have therefore proved that the beam propagation effect shown here in this paper therefore does not depend on the detailed timing of the photon entering the structure. 

\section{Photonic de Haas-van Alphen effect}
In this section, we show a photonic de Haas-van Alphen effect in a lattice with uniform effective magnetic field, which is an exact analogue of the electronic de Haas-van Alphen effect.

We first briefly review the electronic de Haas-van Alphen effect. We consider a solid as described by an electronic band structure $\epsilon(\vec k)$, where $\vec k$ is the Bloch momentum of the electrons. For simplicity, we assume a two dimensional case where the solid and hence the wavevector is restricted to the $x-y$ plane. In the presence of a perpendicular external magnetic field, the motion of the Bloch electrons is described by semiclassical equations \cite{ashcroft}: 
\bqa  \label{req}\frac{d\vec r }{dt}=\vec v_g\equiv\nabla_{\vec k}\epsilon,\\ \label{keq}\frac{d\vec k}{dt}=\vec v_g\times qB \hat{\vec z} \eqa where $q$ is electron charge, and $\vec v_g$ denotes the group velocity. From Eq. (\ref{keq}), we have $d\vec k\cdot \nabla_{\vec k}\epsilon=0$, and thus the momentum satisfies $\epsilon(k_x,k_y)=\epsilon_0$, which means that the trajectory of the electrons in momentum space is a constant energy contour. Integrating Eqs. (\ref{keq}) and (\ref{req}), we can relate the trajectories in the real and momentum spaces: \bqa k_{x}(t)-k_{x}(t=0)=qB[y(t)-y(t=0)],\\ k_{y}(t)-k_{y}(t=0)=-qB[x(t)-x(t=0)].\eqa  As a result, the trajectory of electrons in real space is 
\begin{widetext}
\be\label{cont} \epsilon\big(qB[y(t)-y(t=0)]+k_{x}(t=0), -qB[x(t)-x(t=0)]+k_{y}(t=0)\big)=\epsilon_0. \ee 
\end{widetext}
We therefore see that the trajectory in real space has the same shape as the constant energy contour. 

Having reviewed the electronic case we now consider the corresponding photonic case. In the configuration as shown in Fig. \ref{lattice}a, photons are subject to a uniform effective magnetic field $B_{\textrm{eff}}=\frac{\phi}{a^2}$. Therefore, the equation of motion for photons in the photonic resonator lattice with uniform effective magnetic field here is similar to that of electrons (Eq. (\ref{keq}) and (\ref{req})), with $qB$ replaced by $B_{\textrm{eff}}$. With rotating wave approximation, we can use the band structure $\epsilon_{\rm rwa}$ of $H_{\rm rwa}$ (Eq. \eqref{rwa1}) to represent the Floquet band structure, and thus $\epsilon(k_x,k_y)=V(\textrm{cos}(ak_x)+\textrm{cos}(ak_y))$. The constant energy contours are shown in Fig. \ref{lattice}b. Applying Eq. (\ref{cont}) to this case, the trajectory of photons in real space in the presence of an effective magnetic field is given by 
\begin{widetext}
\be \epsilon\big(B_{\rm eff}[y(t)-y(t=0)]+k_{x}(t=0), -B_{\rm eff}[x(t)-x(t=0)]+k_{y}(t=0)\big)=\epsilon_0. \ee
\end{widetext}
Thus the photon trajectory in real space has the same shape as the constant energy contour where the initial momentum of the photon beam locates. Moreover, unlike the electronic case, where the de Haas-van Alphen effect only probes the electron on the Fermi surface, in the photonic case here, one can map out the entire band structure by simply changing the photon frequency. 

\begin{figure}[H]
\centering\epsfig{file=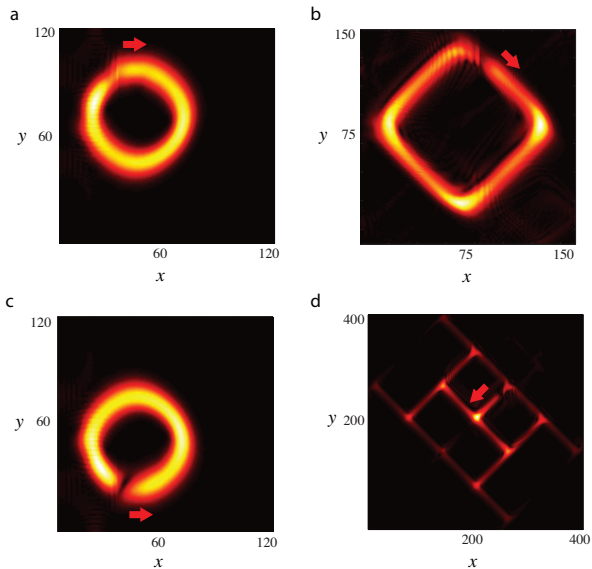,clip=1,width=0.9\linewidth,angle=0}\vspace{0.3cm}\hspace{0.3cm}
\caption{(Color online). Beam trajectories for different initial momenta. The unit of axes is $a$. The initial momenta are: (\textbf{a}) $\vec k_1=-0.41(\pi/a)\hat{\vec x}$, (\textbf{b}) $\vec k_2=-0.48(\pi/a)\hat{\vec x}+0.48(\pi/a)\hat{\vec y}$, (\textbf{c}) $\vec k_3=-0.59(\pi/a)\hat{\vec x}+(\pi/a)\hat{\vec y}$, (\textbf{d}) $\vec k_4=0.5(\pi/a)\hat{\vec x}+0.5(\pi/a)\hat{\vec y}$ as labeled in Fig. \ref{lattice}b. The width $w$ of the source is $\sqrt{50}a$. Red arrows indicate the initial propagation direction. For (\textbf{d}), the beam will eventually trace out a large square grid. Here we show only part of such a grid that has been traced out in the duration of a finite-time simulation. }
\label{trajectory}
\end{figure}

We now numerically demonstrate the photonic de Haas-van Alphen effect. We choose $\phi=0.05$ in the configuration shown in Fig. \ref{lattice}a. We study four different cases with four different initial momenta $\vec k(t=0)$ as indicated in Fig. \ref{lattice}b. The corresponding four trajectories are shown in Fig. \ref{trajectory}.

Case 1: $\vec k(t=0) =-0.41(\pi/a)\hat{\vec x}$ ($\vec k_1$ in Fig. \ref{lattice}b). The initial momentum is near $\Gamma$ point at the center of the first Brillouin zone, where the constant frequency contour is a circle. The corresponding real space trajectory in the presence of the effective magnetic field is indeed a circle (Fig. \ref{trajectory}a). This is equivalent to the demonstration of a Lorentz force for photons as shown in Ref. \cite{our2}.

Case 2: $\vec k(t=0) =-0.48(\pi/a)\hat{\vec x}+0.48(\pi/a)\hat{\vec y}$ ($\vec k_2$ in Fig. \ref{lattice}b). The initial momentum is close to the ridge connecting two neighboring $X$ points, where the constant frequency contour is close to a square. The corresponding real space trajectory is now square-like (Fig. \ref{trajectory}a). 

Case 3: $\vec k(t=0) =-0.59(\pi/a)\hat{\vec x}+(\pi/a)\hat{\vec y}$ ($\vec k_3$ in Fig. \ref{lattice}b). The initial momentum is near an $M$ point, where the constant frequency contour is again a circle. The real space trajectory in the presence of the effective magnetic field is also a circle. However, in contrast to the cases of $\vec k_1$ and $\vec k_2$, here the chirality of the beam trajectory is opposite. The chirality of a beam trajectory is defined as the sign of $\frac{d^2\vec r}{dt^2}\cdot (\frac{d\vec r}{dt}\times \hat{\vec z})$. Since the photon trajectory is closed in these cases, the chirality is used to describe whether the photon moves along the trajectory in a clockwise or a counter closewise direction as viewed from the positive $z$-axis. 

The chirality of the beam trajectory depends on the effective photonic mass as derived from the band structure. For our choices of the initial momentum $\vec k_1$ and $\vec k_3$, the energy band can both be approximated by a quadratic formula, $\epsilon(k_x,k_y)\approx \epsilon_0+\frac{|\vec k-\vec k_0|^2}{2m}$, where $m$ is the effective mass. Substitute the quadratic dispersion into Eqs. \eqref{req} and \eqref{keq}, we have 
\bqa
\frac{d^2\vec r}{dt^2}=\frac{B_{\rm eff}}{m}\frac{d\vec r}{dt}\times \hat{\vec z}.
\eqa Thus the chirality of the photon beam trajectory depends on the sign of the effective mass. Since the effective masses of the photon at $\vec k_1$ and $\vec k_3$ have opposite signs, the chirality of the beam trajectory for these two cases are opposite to each other. 

Case 4: $\vec k(t=0) = 0.5(\pi/a)\hat{\vec x}+0.5(\pi/a)\hat{\vec y}$ ($\vec k_4$ in Fig. \ref{lattice}b). The initial momentum is exactly on the straight line connecting two neighboring $X$ points. The real space trajectory in this case is not a closed trajectory. The momentum of the beam starts tracing along the $\hat{\vec x}-\hat{\vec y}$ direction, towards the $X$ point that is located at $(\pi/a,0)$. This direction is clockwise with respect to the $\Gamma$ point at $(0,0)$, while counter clockwise with respect to the $M$ point at $(\pi/a, \pi/a)$, which is consistent with the discussion above regarding the chirality of the beam trajectory.  When the momentum reaches this $X$ point, the beam in real space splits with equal amplitude into two branches perpendicular to the original beam. This splitting process happens whenever the momentum of the beam reaches an $X$ point in the momentum space. The resulting trajectory in real space is a square grid with a unit cell size of $\sqrt{2}\pi a/\phi$, as shown in Fig. \ref{trajectory}d.  

\section{Experimental implementation and Summary}
In Ref. \cite{our2}, we have provided a detailed discussion of the experimental feasibility of achieving an effective gauge field for photons, in either optical frequency range with electro-optic effect, or in the micro-wave frequency range with the use of a mixer. Here, we only focus on those aspects that are specific to the demonstration of the beam propagation effects as considered in this paper. 

In order to discuss the experimental conditions required to observe the photonic de Hass-van Alphen effect, for concreteness we consider only the closed trajectories. To observe such a single round trip, the beam should not be significantly dissipated after completing a closed trajectory. Consider a beam tracing out a circle of radius $k$ in momentum space. The trajectory in real space then has a radius of $ka^2/\phi$. The group velocity of the beam is $Va^2k$ and thus the time for the beam to circulate once is $T=2\pi/(V\phi)$. If we require the loss of the beam to be less than 3 dB after one circulation, then the intrinsic loss rate of the resonator cannot exceed $\frac{\textrm{ln}2}{(2\pi)}V\phi$. The intrinsic loss rate is related to the $Q$ factor of the resonator as $\frac{\omega}{2Q}$. For an operating frequency of $\omega=2\pi\cdot200$ THz, which corresponds to an operating wavelength near 1.5 micron, assuming a coupling constant of $V=10$ GHz, and an effective magnetic field that corresponds to $\phi=0.4$, this requirement sets the $Q$ factor to be greater than $1.4\times 10^6$, which is achievable in the state-of-the-art photonic crystal resonators \cite{noda,tanabe}. 

In summary, we have proposed a photonic de Haas-van Alphen effect using effective magnetic field in a dynamically modulated two-dimensional square photonic resonator lattice. Such novel beam steering can also be similarly achieved in other kinds of lattices, which have different constant energy contours, and thus results in diverse shapes of trajectories. Moreover, with the availability of three dimensional photonic crystals \cite{3dpc} and three-dimensional on-chip integration \cite{3dinte}, it is possible that such an effect can also be realized in three dimensions. 

This work is supported in part by U. S. Air Force Office of Scientific Research grant No. FA9550-09-1-0704, and U. S. National Science Foundation grant No. ECCS-1201914.

\pagebreak

%
%

\end{document}